\documentclass[prb,twocolumn,showpacs]{revtex4}

\usepackage[utf8]{inputenc}
\usepackage{graphicx}
\usepackage{dcolumn}
\usepackage{bm}
\usepackage{amsmath}
\usepackage{amssymb}
\usepackage{times}

\usepackage{xcolor}

\renewcommand{\b}{\ensuremath{\beta}}
\renewcommand{\d}{\ensuremath{\delta}}
\newcommand{\e}{\ensuremath{\epsilon}}
\newcommand{\g}{\ensuremath{\gamma}}
\renewcommand{\i}{\ensuremath{\boldsymbol{i}}}
\newcommand{\G}{\ensuremath{\Gamma}}
\newcommand{\MatW}{\ensuremath{\boldsymbol{W}}}
\newcommand{\MatRel}{\ensuremath{\boldsymbol{\Theta}}}
\newcommand{\VB}{\ensuremath{V_{\text{bias}}}}

\newcommand{\op}[1]{\ensuremath{\Hat{#1}}}
\newcommand{\hop}[1]{\ensuremath{\op{#1}^{\dagger}}}
\newcommand{\lto}{\ensuremath{\leftarrow}}
\newcommand{\W}{\ensuremath{\text{W}}}
\renewcommand{\P}{\ensuremath{\text{P}}}
\newcommand{\PVec}{\ensuremath{\text{\textbf{P}}}}

\newcommand{\bra}[1]{\ensuremath{\langle{#1}\vert}}
\newcommand{\ket}[1]{\ensuremath{\vert{#1}\rangle}}
\newcommand{\abs}[1]{\ensuremath{\mathinner{\lvert#1\rvert}}}
\renewcommand{\vec}[1]{\ensuremath{\boldsymbol{#1}}}
\newcommand{\expect}[1]{\ensuremath{\langle{#1}\rangle}}

\newcommand{\QDD}{\ket{\text{d}}}
\newcommand{\QDg}{\ket{\text{g}}}
\newcommand{\QDe}{\ket{\text{e}}}

\DeclareMathOperator{\Tr}{Tr}

\usepackage%
[%
	pdftex,%
	colorlinks=true,%
	linkcolor=blue,%
	citecolor=blue,%
	urlcolor=blue,%
	letterpaper=true%
]{hyperref}

\begin{document}

\title{Coulomb-Blocked Transport Through a Quantum Dot with Spin-Split Level:\\
       Increase of Differential Conductance Peaks by Spin Relaxation}

\author{Daniel Becker}
 \email{dbecker@physnet.uni-hamburg.de}
\author{Daniela Pfannkuche}
\affiliation{
	I. Institute for Theoretical Physics, University of Hamburg, D-20355 Hamburg, Germany
}

\date{\today}

\begin{abstract}
Non-equilibrium transport through a quantum dot with one spin-split single-particle level is studied in the cotunneling regime at low temperatures. The Coulomb diamond can be subdivided into parts differing in at least one of two respects: what kind of tunneling processes (i) determine the single-particle occupations and (ii) mainly contribute to the current. No finite systematic perturbation expansion of the occupations and the current can be found that is valid within the entire Coulomb diamond. We therefore construct a non-systematic solution, which is physically correct and perturbative in the whole cotunneling regime, while smoothly crossing-over between the different regions. With this solution the impact of an intrinsic spin-flip relaxation on the transport is investigated. We focus on peaks in the differential conductance that mark the onset of cotunneling-mediated sequential transport. It is shown that these peaks are maximally pronounced at a relaxation roughly as fast as sequential tunneling. The approach as well as the presented results can be generalized to quantum dots with few levels.
\end{abstract}


\maketitle

\section{Introduction} \label{sec:Intro}

\vspace{-0.1mm}Coherence effects, whose signatures can be seen in $\text{(spin-)}$ electronic transport through low dimensional nanoscopic structures like quantum dots\cite{PhysRevA.57.120, PhysRevB.61.12639, 0034-4885-64-6-201}, provide insight into fundamental aspects of quantum mechanics and have important applications in vital fields of research such as spintronics, quantum computing, and data storage. \cite{2001Sci...294.1488W, 2007NatPh...3..153A} For the occurrence of these effects spin-flip relaxation is widely considered as a limiting factor and therefore usually sought to be as small as possible. In experiments spin-flip relaxation times $T_1$ ranging from $\mu$s\cite{2002Natur.419..278F, PhysRevLett.91.196802} to ms\cite{2004Natur.430..431E, 2004Natur.432...81K} have been observed displaying dependences both on the sort of quantum dot and on the parameters of the experimental setup like temperature and magnetic field \cite{PhysRevB.61.12639, golovach:016601, 2006cond.mat..6782C, 0034-4885-64-6-201}. Accordingly, the relaxation rates can be experimentally adjusted in a wide range either by means of tuning of external parameters or by suitably tailoring the quantum dot itself. Recently, spin-flip times of even several hundred milliseconds were measured in $n$-doped (In, Ga)As/GaAs quantum dots charged with spin-polarized electrons at low magnetic field and temperature. \cite{sigrist:036804} For transport through few-electron quantum dots in the presence of intrinsic spin relaxation, which is discussed in this paper, microscopic mechanisms like phonon-induced spin decay due to spin-orbit or hyperfine interaction have been investigated theoretically (see, e.g., Refs. \onlinecite{golovach:016601, PhysRevB.61.12639, lehmann:045328, 2006cond.mat..6782C, 2007arXiv0712.2376V, PhysRevB.64.195306}).

Though relaxation mostly acts destructively on coherent electron dynamics, it can also, however, considerably pronounce their effect, as \textsc{Weymann} \textit{et al.} \cite{weymann:205309} show for the case of Coulomb-blocked transport through a single-level quantum dot (SLQD) coupled to ferromagnetic leads with antiparallel magnetization. The zero-bias anomaly of the differential conductance and the conductance step at the onset of inelastic cotunneling are increased by a slow spin relaxation for a spin-degenerate and a spin-split dot level, respectively. 

We will show in this paper that a similar, strongly pronounced effect should be observable even when the leads are non-magnetic. In the considered case a small relaxation that is roughly as large as the tunnel coupling maximizes peaks of the differential conductance, which mark the onset of cotunneling-mediated sequential transport. This effect is associated with sequential tunneling out of an excited single-particle state. Within a SLQD model, the dot's level has to be spin-split. For few-electron GaAs/$\text{Al}_{0.3}\text{Ga}_{0.7}\text{As}$ quantum dots with non-degenerate orbital levels these signatures of the single-particle spectrum have been intensely studied both experimentally and theoretically by \textsc{Schleser} \textit{et al.} \cite{schleser:206805}, whereby new insight was provided into the interplay between sequential and cotunneling in the Coulomb blockade regime. In the present paper we investigate this interplay in further detail. Since said signatures appear close to resonance with single-particle transitions, we have to base our calculations on the non-equilibrium Keldysh formalism rather than second-order perturbation theory. \cite{golovach:245327, Tews:2004, schleser:206805} It is shown that in order to obtain physically correct, perturbative results for the entire cotunneling regime, one has to construct non-systematic rate equations similar to those proposed in Ref. \onlinecite{weymann:115334}. In the latter equations we identify terms that cannot belong to the second-order perturbation expansion. By omitting these terms, one ensures that for the considered system the occupation probabilities are well-defined everywhere in the Coulomb blockade regime. As in Ref. \onlinecite{weymann:205309} we treat the effect of relaxation phenomenologically, describing the intrinsic spin-flip processes by an effective rate $\theta$. Thus no particular mechanism has been specified.

The paper is structured as follows. In Sec. \ref{sec:Model} we introduce the model system and explain restrictions on the system parameters. The diagrammatic transport theory and the derivation of transport equations are sketched out in Sec. \ref{sec:Theory}. Results are presented and discussed in Sec. \ref{sec:Results} and followed by a summary in Sec. \ref{sec:Summary}.

\section{Model} \label{sec:Model}

We consider a model system consisting of a SLQD, which is coupled to two metallic leads (L and R) by identical tunneling barriers, so that a dc bias voltage $\VB$, symmetrically applied between both reservoirs, causes a tunneling current through the dot. An additional capacitatively coupled gate electrode allows to adjust the electrostatic potential $\Phi_D$ in the dot by applying a gate voltage. Such a system can be represented by the Anderson-type Hamiltonian $\op{H} = \op{H}_D + \op{H}_L + \op{H}_R + \op{H}_T$ with the quantum dot part $\op{H}_D$, the Hamiltonians $\op{H}_{\text{L}}$ and $\op{H}_{\text{R}}$ of the left and right lead, respectively, and the tunneling operator $\op{H}_T$, describing the coupling between the dot and the leads. We assume that the spin degeneracy of the two single-electron dot states is lifted (e.g., by a Zeeman-field), leading to an energy difference of $\Delta$. Then the dot Hamiltonian can be written as $\op{H}_D = \e\, \hop{a}_{\text{g}} \op{a}_{\text{g}} + (\e + \Delta)\, \hop{a}_{\text{e}} \op{a}_{\text{e}}+ U\, \hop{a}_{\text{e}}\op{a}_{\text{e}}\hop{a}_{\text{g}}\op{a}_{\text{g}}$.
Here the index g (e) denotes the spin of the single-electron ground state $\QDg$ (excited state $\QDe$) and $\hop{a}_\sigma$ ($\op{a}_\sigma$) with $\sigma \in \{ \text{g}, \text{e} \}$ creates (annihilates) an electron with spin $\sigma$ and energy $\e_{\sigma} = \e + \d_{\sigma, \text{e}} \Delta$ when acting on the empty dot state $\ket{0}$ ($\d$ is the Kronecker delta). $U$ is the Coulomb-energy of the doubly occupied state $\QDD$. The leads play the role of macroscopic reservoirs and are described as free electron gases with Hamiltonian $\op{H}_r = \sum_{\vec{k},\sigma} \e_{\vec{k},r}\, \hop{c}_{\vec{k},\sigma,r} \op{c}_{\vec{k},\sigma,r}$, where $r \in \{\text{L},\text{R} \}$ refers to the lead; $\vec{k}$ is the wave vector of an electron in reservoir $r$, $\sigma$ is its spin and $\e_{\vec{k},r}$ its energy. The $\hop{c}_{\vec{k},\sigma,r}$ ($\op{c}_{\vec{k},\sigma,r}$) are the corresponding creation (annihilation) operators. Due to the applied bias voltage, $\mu_r = (-1)^{\d_{r,\text{L}}} e \VB / 2$---with $e > 0$ being the elementary charge---gives the electrochemical potential of reservoir $r$. The coupling between the leads and the dot is described by $\op{H}_T = \sum_{\vec{k},\sigma, r} \bigl(\g\, \hop{a}_{\sigma} \op{c}_{\vec{k},\sigma,r} + h.c.\bigr)$, where the first (second) term on the rhs describes tunneling into (out of) the dot with the complex tunneling parameter $\g$ ($\g^*$), which is assumed to be independent of wave vector and spin of a tunneling electron as well as the reservoir out of which (into which) it tunnels. With the constant density of states $\rho$ of the reservoirs the coupling can be characterized by the positive scalar parameter $\G := \abs{\g}^2 \rho$ alone.\cite{UnconventionalCoupling} The stationary tunneling current $I$ is the expectation value of the current operator $\op{I} := \op{I}_{\text{L}}$, where $\op{I}_r = -\i (e / \hbar) \sum_{\vec{k},\sigma} \bigl( \g\, \hop{a}_{\sigma} \op{c}_{\vec{k},\sigma,r} - h. c. \bigr)$ \cite{ChargeConservation}. 

We demand that the reservoirs stay in equilibrium even when coupled to the SLQD. For a perturbative calculation of the occupation probabilities and the current up to second order in the tunnel coupling, $\G$ has to be very small compared to the dot energies $\e$ and $U$. Coulomb blockade of sequential transport is possible, if the thermal energy is very much smaller than $U$, i.e., $\b^{-1} \equiv k_{\text{B}}T \ll U$ with $T$ being the temperature and $k_{\text{B}}$ Boltzmann's constant. We restrict our study to parameter sets with $\b \G \ll 1$, which is a necessary condition for physical behavior of the second-order perturbation expansion in $\G$ once the electro-chemical potential of a reservoir is close to resonance with the energy of a single-charge excitation. \cite{thielmann:146806} 
Furthermore, to be able to see transport signatures of the excited state within the Coulomb blockade regime, the Zeeman-splitting $\Delta$ must not be very much smaller than $U$ but roughly of the same order of magnitude (though not larger than $U/2$). This implies $\b \Delta \gg 1$. For the particular parameter set $(\G = 4.5 \times 10^{-3} k_{\text{B}}T, \Delta=45 k_{\text{B}}T, U = 225 k_{\text{B}}T)$ we use throughout the following discussions, this requirement may be difficult to meet for quantum dots made of GaAs or Si and magnetic fields available in laboratories. On the other hand, the presented perturbative framework can be applied to systems with $\G$ that is up to 10 times larger and with $U, \Delta$ that are 10 times smaller, while yielding the same qualitative results. In practice, since it is purely of mathematical origin, the criterion $\b \G \ll 1$ is not experimentally relevant and imposes no restriction on the physics underlying the transports effects, we present here. Therefore, the results of our approach can also be applied to experiments on GaAs or Si quantum dots in which the split-exceeded level can be resolved in transport spectroscopy. \cite{2007RvMP...79.1217H} InAs nanowire quantum dots, however, have an effective g Factor between 8 and 9\cite{fasth:266801} and an effective mass of $m^* = 0.02 m_e$ ($m_e$ is the mass of a free electron).\cite{pfund:036801} In experiments with these dots an adequately large spin-splitting should be feasible for Zeeman fields in the range of 1 to 10 T, even for the parameter set we use here. 

We would also like to emphasize in this context, that the presented approach is not restricted to quantum dots with one spin-split single-particle level but can in the same way be employed for dots with two non-degenerate spinless orbitals. For very similar systems (few-electron GaAs/AlGaAs quantum dots) the discussed conductance peaks were seen in low magnetic fields. \cite{schleser:206805}

\section{Master equations} \label{sec:Theory}

For our calculations we use the real-time diagrammatic technique developed by Schoeller et al.\cite{PhysRevB.50.18436} It is based on the Keldysh formalism and allows to represent a dynamical, non-equilibrium property of the model system by a formaly exact, infinite perturbation expansion with small parameter $\G$. From such an expression one can obtain a systematically expanded quantity up to a finite order in the coupling.\cite{PhysRevLett.76.1715, PhysRevB.54.16820, PhysRevLett.78.4482, PhysRevB.68.115105, thielmann:045341, braggio:026805} In order to construct perturbative solutions for the occupation probabilities and the tunneling current, we first assume that intrinsic relaxation is absent. To compute the time-dependent statistical expectation value $\Tr \bigl(\op{\rho}(t) \op{I}^r \bigr)$ of the current operator, we have to calculate the density matrix $\op{\rho}(t)$, which contains the complete system dynamics. Since the reservoirs are assumed to stay in equilibrium at all times, the density matrix's reservoir degrees of freedom can be integrated out using Wick's theorem. This yields the \emph{reduced density matrix} $\op{\rho}_D (t)$, which depends only on the dot degrees of freedom. Via an adiabatic switching between times $t_0$ and $t$ the initial state of the isolated dot, represented by $\op{\rho}^0_D \equiv \op{\rho}_D (t_0)$, is connected to the reduced density matrix of the coupled system $\op{\rho}_D (t)$. This relation is expressed by equation 
\begin{equation}\label{eqn:KineticEquation1}
	\op{\rho}_D (t) = \op{\Pi} (t, t_0) \op{\rho}^0_D, 
\end{equation}
where $\op{\Pi} (t,t')$ is a time evolution operator describing propagation of the reduced density matrix between $t'$ and $t$. The propagator $\op{\Pi} (t,t')$ can be represented as an infinite sum of diagrams on the Keldysh contour, each of which is decomposable into parts $\op{\Pi}^0$ corresponding to propagation that is not influenced by the reservoirs and irreducible self-energy parts that describe coherent dynamics governed by the tunnel coupling and allow the dot to change its state \cite{PhysRevB.50.18436}. With the operator $\op{\Sigma}$, which consists of all irreducible diagrams, a Dyson equation for $\op{\Pi}$ can be set up leading to the \emph{kinetic equation}
\begin{equation}\label{eqn:KineticEquation2}
\begin{split}
	\op{\rho}_D (t) &= \op{\Pi}^0 (t,t_0) \op{\rho}^0_D \\
			&+ \int_{t_0}^t d t_2 \int_{t_0}^{t_2} d t_1 
			   \op{\Pi}^0 (t,t_2) \op{\Sigma} (t_2,t_1) \op{\rho}_D (t_1),
\end{split}
\end{equation}
when plugged into (\ref{eqn:KineticEquation1}). In the limit of $t_0 \to -\infty$ and vanishing adiabatic switching, the derivative of Eq. (\ref{eqn:KineticEquation2}) with respect to $t$ becomes a self-consistent conditional equation for the stationary reduced density matrix $\op{\rho}_D^{\text{st}}$, provided that the SLQD will eventually forget its initial state $\op{\rho}^0_D$ due to the interaction with the macroscopic reservoirs \cite{StationaryLimit}. Since we assume diagonality of the initial density matrix $\op{\rho}^0_D$, which implicates diagonality of $\op{\rho}_D^{\text{st}}$ \cite{DiagramStructure}, it is convenient to replace the latter by the vector $\PVec$ of the stationary probabilities $\P_{\phi} = \bra{\phi} \op{\rho}_D^{\text{st}} \ket{\phi}$ for the dot to be in state $\ket{\phi}$ with $\phi \in \{ 0, \text{g}, \text{e}, \text{d} \}$. We then replace the tensor operator $\op{\Sigma}$ with the matrix $\MatW$, whose elements
\begin{equation} \label{eqn:WMatrix}
	\W_{\phi'\lto\phi} := \int_{-\infty}^{0} d t' \Sigma_{\phi'\lto\phi}^{\phi'\lto\phi} (0,t')
\end{equation}
are interpreted as stationary rates of quantum dot transitions from state $\ket{\phi}$ to state $\ket{\phi'}$. Since the total probability has to be conserved, the $4\times 4$ Matrix $\MatW$ has a rank of three. Therefore the resulting self-consistent \emph{rate equation}
\begin{equation} \label{eqn:RateEquation}
	\MatW \PVec = \boldsymbol{0}
\end{equation}
has non-trivial solutions and, together with the normalization condition $\sum_{\phi} \P_{\phi} = 1$, uniquely determines $\PVec$ as a vector of probabilities.\cite{thielmann:045341, weymann:115334} A similar equation for the current $I^r$ out of reservoir $r$ into the dot can be formed, if we introduce an operator $\op{\Sigma}^r$, which consists of all irreducible diagrams of $\op{\Sigma}$, each having its last internal vertex replaced by an external vertex stemming from $\op{I}^r(t=0)$. With a matrix $\MatW^r$, defined in analogy to (\ref{eqn:WMatrix}), we get
\begin{equation} \label{eqn:CurrentEquation}
	I = -e \sum_{\phi} \bigl( \MatW^{\text{L}} \PVec \bigr)_{\phi}
	  =  e \sum_{\phi} \bigl( \MatW^{\text{R}} \PVec \bigr)_{\phi}.
\end{equation}

Each of the Eqs. (\ref{eqn:RateEquation}) and (\ref{eqn:CurrentEquation}) yields an infinite system of coupled equations, if we express $\MatW$, $\MatW^{\text{L}}$, $\PVec$, and $I$ as expansions in $\G$ and sort all terms by order. The $n^{\text{th}}$-order occupation vector and current are given by
\begin{equation}\label{eqn:NOrderRateEq}
\begin{split}
	\boldsymbol{0} &= \sum_{l=1}^n \MatW^{(l)}\PVec^{(n-l)} \\
	I^{(n)} &= -e \sum_{\phi} 
			\sum_{l=1}^n \bigl( \MatW^{\text{L}(l)} \PVec^{(n-l)} \bigr)_{\phi}.
\end{split}
\end{equation}
With the terms $\MatW^{(n)}$ and $\MatW^{r (n)}$ we identify those parts of $\MatW$ and $\MatW^r$, respectively, that are represented by irreducible diagrams with exactly $n$ tunneling lines. Each of these lines connects two vertices on the Keldysh contour and represents the wick contraction of the corresponding reservoir operators. The ascending orders of $\PVec$ and $I$ are then calculated iteratively, starting with $\PVec^{(0)}$ and $I^{(1)}$, where the $\PVec^{(n)}$ are normalized according to $\sum_{\phi} \PVec^{(n)}_{\phi} = \d_{n,0}$. The first- and second-order equations describe transport caused by sequential tunneling and cotunneling processes, respectively. The curve of the sequential current against the bias voltage resembles a staircase with thermally broadened steps formed at bias values appropriate for single-charge excitations. Cotunneling further broadens these steps \cite{thielmann:146806} and dominates the transport behavior within the \emph{Coulomb blockade regime} (or cotunneling regime), where the gate voltage is tuned to charge the SLQD with one electron, while the bias is too small to doubly occupy or to empty the dot. \cite{1989PhLA..140..251A, PhysRevLett.65.2446, PhysRevLett.65.3037, golovach:245327}

\begin{figure}[htbp]

	\includegraphics[scale=1]{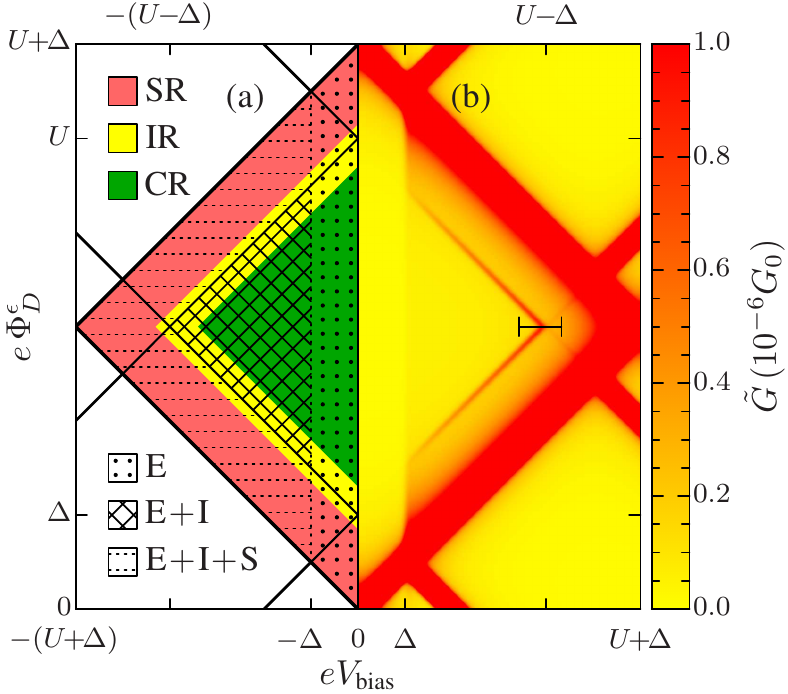}

	\caption{\label{fig:ChargingDiagram}(Color online). (a) Schematic picture of the diamond-shaped
	cotunneling regime showing its subdivision into areas with different possible tunneling
	processes (hatched areas) as well as into core (CR), shell (SR), and intermediate region (IR), 
	in which the occupations $\P_{\text{g}}$ and $\P_{\text{e}}$ are determined either by
	cotunneling, sequential tunneling, or both (colored (shaded) areas). E stands for elastic 
	and I for inelastic cotunneling, S is for sequential tunneling. (b) Calculated charging diagram 
	(based on Eqs. (\ref{eqn:ApproxRateEqCurr}) and (\ref{eqn:ApproximateRateEq})) of the 
	cotunneling regime with parameters $\Delta = 45 k_{\text{B}} T$, $U = 225 k_{\text{B}} T$, 
	$\G = 4.5 \times 10^{-3} k_{\text{B}} T$ and $\theta = \G / 2$. High values of 
	the differential conductance $\tilde{G} = d \tilde{I} / d \VB$ (in units of $G_0 =
	\b\G\,e^2/\hbar$) outside and at the border of the diamond are clipped by the color scale 
	(values near the border and in the exterior: see text). The short horizontal line corresponds 
	to the range of bias values in Fig. \ref{fig:DiffCondPeaks}. Both (a) and (b) can be extended 
	to regions with opposite sign of $e \VB$ by reflection with respect to the $e \VB = 0$ axis.}
\end{figure}

Depending on its strength $\G$, the coupling of a microscopic system like a quantum dot to macroscopic reservoirs will modify the dot's behavior slightly ($\G \ll \e,U$) or drastically ($\G \gg \e,U$). Even in the first case \cite{ThesisKoenig} and for incoherent sequential transport \cite{wunsch:205319}, the coupling may be too strong to assume that the dot propagates as if isolated between two tunneling events. In such cases and all the more when transport of highly correlated electrons is considered \cite{PhysRevLett.88.226601, sandalov:075315}, it may still be possible to represent the dot as isolated but with \emph{renormalized} instead of the \emph{bare} system parameters. Strictly speaking, this is also true for the system studied here and correctly accounted for by the diagrammatic technique. \cite{ThesisKoenig} However, since the difference between renormalized and bare energies scales with $\G$, for the regime we investigate ($\G \ll \b^{-1} \ll \e,\Delta, U$) it is so small, that the effect of the renormalization on the discussed transport phenomena is virtually unobservable. This can be seen, for example, in Fig. \ref{fig:ChargingDiagram}(b), where a light vertical shade within the Coulomb blockade regime indicates the onset of inelastic cotunneling as soon as $\abs{e \VB}$ equals the \emph{renormalized} excitation energy (see below), while on the $e \VB$ axis the position of the \emph{bare} spin-splitting $\Delta$ is marked. Obviously, both positions do not considerably deviate from each other. Hence, throughout the following discussions, we do not distinguish between the bare and renormalized quantities, although all statements are strictly valid only for the latter ones.

Fig. \ref{fig:ChargingDiagram}(a) schematically shows the (sub-)structure of the cotunneling regime plotted against $e \VB$ and the dot potential energy. For convenience the latter is given relative to the energy that is needed to charge the dot with one electron in the ground state: $e\,\Phi_D^{\e} := e\,\Phi_D - \e$. On the one hand the diamond-shaped cotunneling regime (Coulomb diamond) breaks down into three regions differing with respect to the kind of tunneling processes that predominantly determine the occupation of the single-particle states. As we discuss below $\P_{\text{g}}$ and $\P_{\text{e}}$ are given by sequential tunneling in the red colored (medium gray) \emph{shell region} (SR), by cotunneling in the green (dark gray) \emph{core region} (CR) and by a mixture of both in the yellow (light gray) \emph{intermediate region} (IR). 

On the other hand one can distinguish three sub-regimes of the Coulomb diamond with different current-driving tunneling processes. The quantum dot is in state $\QDg$ ($\P_{\text{g}} \approx 1$) for $\abs{e \VB} < \Delta$ (dotted area) and the small finite current is maintained just by energy-conserving \emph{elastic} cotunneling (E) through virtual states $\ket{0}$ and $\QDD$ (elastic regime). Once $\abs{e \VB}$ exceeds $\Delta$, \emph{inelastic} cotunneling processes can excite the dot into state $\QDe$ while transferring energy from the reservoirs into the SLQD.\cite{PhysRevLett.86.878, schleser:206805} Since each of these processes effectively carries one electron through the dot, they cause an additional electron flow (+I). When passing from the core to the shell part of this inelastic cotunneling regime---the corresponding areas are hatched with crossed lines and dashed horizontal lines, respectively---cotunneling-mediated \emph{sequential} tunneling out of the excited state sets on and further increases the current (+S). \cite{schleser:206805}

Before we can study the transport in vicinity of the excited state resonances, it is necessary to modify the rate equations (\ref{eqn:NOrderRateEq}), as they prove to be unsuitable to describe the occupations and current in the intermediate region. As \textsc{Weymann} \textit{et al.} show in Ref. \onlinecite{weymann:115334} it is due to the breakup of the cotunneling regime into core, shell, and intermediate region that no systematic second-order expansion of $\PVec$ or $I$ exists, which is valid within the entire regime. This can be explained as follows.

In the shell region only those sequential transitions are energetically forbidden---the corresponding rates being exponentially small---that carry the dot out of the ground state $\QDg$. Hence, after a finite time of propagation, the dot inevitably gets trapped in $\QDg$ and thereby forgets its initial state. So the stationary occupations are essentially determined by sequential tunneling alone. Even with all rates $\W_{\phi'\lto\text{g}}^{(1)}$ set to zero, the matrix $\MatW^{(1)}$ still has a rank of three and by Eq. (\ref{eqn:RateEquation}) all $P_{\phi}^{(0)}$ are fixed except for normalization. 

In the core region sequential transitions out of both single-particle states are forbidden. Then classically the dot can get trapped either in the ground or in the excited state, so that the single-particle occupations depend on the initial dot state. Consequently, they are no longer determined by the lowest- but by the second-order rate equation, i.e., they are essentially given by cotunneling. This becomes manifest in the structure of $\MatW^{(1)}$, which has a rank of two, when all rates $W_{\phi'\lto\text{g},\text{e}}^{(1)}$ are set to zero. 

Between shell and core lies the intermediate region, where the system continuously changes between classical and cotunneling-dominated occupation, respectively. But as well as no matrix $\MatW^{(1)}$ can be constructed that continuously changes its rank, no single rate equation exists that both determines the systematic second-order expansion of $\PVec$ in terms of $\MatW^{(1)}$, $\MatW^{(2)}$ and is valid within all three regions simultaneously. 

Alternatively, we seek second-order approximations of $\PVec$ and $I$, which are valid in the cotunneling regime and perturbative in the sense that they deviate from the systematic expansions at most by terms quadratic and cubic in $\G$, respectively. With the normalized solution $\PVec'$ of Eq. (\ref{eqn:RateEquation}), in which $\MatW$ is replaced by the sum of the lowest two orders $\MatW^{(1+2)}$, \textsc{Weymann} \emph{et al.} present an example for an approximation that is perturbative even for arbitrary values of $e \VB$ and $e\,\Phi_D^{\e}$.\cite{weymann:115334} Unfortunately, for our system $\PVec'$ isn't well-defined within the entire core region, where the component $\P'_{\text{e}}$ becomes negative when $\abs{e \VB} < \Delta$ (green (dark gray), dotted area in Fig. \ref{fig:ChargingDiagram}(a)). To resolve this problem we take into account that on the rhs of the second-order equation\cite{SpecialCase}
\begin{equation} \label{eqn:SecondOrderCurrent}
	 \MatW^{(1)} \PVec^{(1)} = -\MatW^{(2)} \PVec^{(0)}
\end{equation}
the first-order probabilities $\P^{(0)}_{0,\text{d}}$, which are exponentially small within the cotunneling regime, are multiplied with the rates $W^{(2)}_{\phi'\lto 0,\text{d}}$. Hence, these rates drop out of Eq. (\ref{eqn:SecondOrderCurrent}) and its rhs reduces to a vector $\vec{V}$ with components $V_{\phi'} = -\sum_{\phi}(\d_{\phi,\text{g}}+\d_{\phi,\text{e}}) W_{\phi'\lto\phi}^{(2)}$. As a consequence the rates $W^{(2)}_{\phi'\lto 0,\text{d}}$ don't contribute to systematic expansion orders given solely in terms of $\MatW^{(1)}$ and $\MatW^{(2)}$, and all contributions to $\PVec'$, they are contained in, are unsystematic and should be omitted. With regard to the approximation of the current the same is true for terms containing the rates $W^{\text{L} (2)}_{\phi'\lto 0,\text{d}}$. By dropping the unsystematic terms we arrive at
\begin{align} 
	\boldsymbol{0} &= \bigl( \MatW^{(1)} + \tilde{\MatW}^{(2)}\bigr) \tilde{\PVec},
	\label{eqn:ApproxRateEqOccu}\\
	\tilde{I} &= -e \sum_{\phi} \bigl( (\MatW^{\text{L}(1)}
			 +\tilde{\MatW}^{\text{L}(2)} )	\tilde{\PVec} \bigr)_{\phi},
	\label{eqn:ApproxRateEqCurr}
\end{align}
where $\tilde{W}^{(2)}_{\phi'\lto\phi} = (\d_{\phi,\text{g}}+\d_{\phi,\text{e}}) W_{\phi'\lto\phi}^{(2)}$.

Finally, without specifying a particular spin-flip mechanism, we include relaxation via an effective Hamiltonian
\begin{equation} \label{eqn:EffRelaxHamiltonian}
	\op{H}_{\text{rel}} = \sum_{q} \bigl( \tau \hop{a}_{g} \hop{b}_{q} \op{a}_{e} + h.c. \bigr),
\end{equation}
which describes the coupling of the dot electrons to a bath of free particles with temperature $T$ and $\op{H}_{\text{bath}} = \sum_q \e_q \hop{b}_q \op{b}_q$\cite{QuantumNumbers}. This coupling is characterized by the single complex parameter $\tau$, giving the amplitude for a spin-flip process from $\QDe$ to $\QDg$. We assume, that the relaxation processes are completely incoherent to the electron tunneling and include only the first order of the perturbation expansion with respect to $\op{H}_{\text{rel}}$. Then, in the diagrammatic representation, the self-energy operator up to second order becomes the sum of all irreducible diagrams that have \emph{either} up to two tunneling lines $\bigl(\op{\Sigma}^{(1,2)}\bigr)$ \emph{or} exactly one relaxation line $\bigl( \op{\Sigma}_{\text{rel}}^{(1)} \bigr)$, which represents a wick contraction of bath operators. The latter operator gives rise to an additional matrix term $\MatRel$ in the master equation, whose matrix elements are defined in analogy to Eq. (\ref{eqn:WMatrix}). Hence, we get the rate equation
\begin{equation} \label{eqn:ApproximateRateEq}
	\boldsymbol{0} = \bigl( \MatW^{(1)} + \tilde{\MatW}^{(2)} + \MatRel \bigr) \tilde{\PVec}
\end{equation}
for a relaxation-dependent approximation $\tilde{\PVec}$. Evaluation of the relaxation diagrams then yields the rates 
\begin{equation} \label{eqn:RelaxationRates}
	\begin{split}
		\Theta_{\text{g}\lto\text{e}} = -\Theta_{\text{e}\lto\text{e}}
			&= \frac{2 \pi \abs{\tau}^2}{\hbar} \int d \e_q 
			\expect{\op{b}_{q} \hop{b}_{q}}_{\text{b}} \rho_{\text{b}} (\e_q)
			\d(\e_q - \Delta) \\
		\Theta_{\text{e}\lto\text{g}} = -\Theta_{\text{g}\lto\text{g}}
			&= \frac{2 \pi \abs{\tau}^2}{\hbar} \int d \e_q 
			\expect{\hop{b}_{q} \op{b}_{q}}_{\text{b}} \rho_{\text{b}} (\e_q)
			\d(\e_q - \Delta),
	\end{split}
\end{equation}
where $\rho_{\text{b}} (\e_q)$ gives the density of states in the bath at energy $\e_q$ and $\expect{\cdot}_{\text{b}}$ denotes the expectation value with respect to the bath degrees of freedom. Alternatively, these spin-flip rates can be calculated using standard time-dependent perturbation theory and Fermi's Golden rule (see, e.g., Ref. \onlinecite{ThesisWeinmann}). The first equalities in the Eqs. (\ref{eqn:RelaxationRates}) express the conservation of the total probability, which in the diagrammatic approach is fulfilled by construction. Since we assume that the spin-splitting is large compared to the temperature ($\Delta \b \gg 1$), the relaxation rates are approximately given by $\Theta_{\phi'\lto\phi} = \d_{\phi,\text{e}} (\d_{\phi',\text{g}}-\d_{\phi',\text{e}}) \theta / \hbar$ with $\theta = 2 \pi \abs{\tau}^2 \rho_{\text{b}} (\Delta)$ both for a fermionic and a bosonic bath (as long as $\abs{\mu_{\text{bath}}} \ll \Delta$).\cite{RelaxDefWeymann} $\tilde{\PVec}$ as well as $\tilde{I}$ are well-defined and perturbative within the cotunneling regime and seamlessly link in the intermediate region the systematic expansions that are only valid either in the core or shell. We also note that, because $\op{H}_{\text{rel}}$ and $\op{H}_r$ commute, the relaxation does not contribute (directly) to the current $\tilde{I}$, that is to say, Eq. (\ref{eqn:ApproxRateEqCurr}) remains valid without modification.

\begin{figure}[htbp]

	\includegraphics[scale=1]{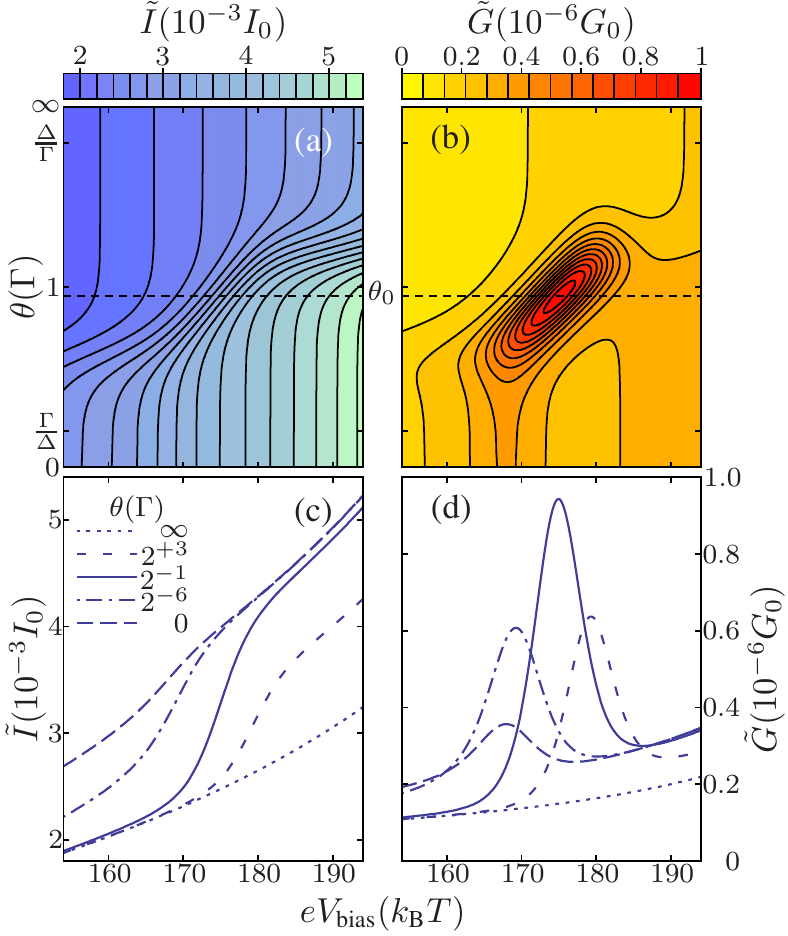}

	\caption{\label{fig:DiffCondPeaks} (Color online). Current $\tilde{I}$ (in units of 
	$I_0 = e \G / \hbar$) and differential conductance $\tilde{G}$ versus bias $e \VB$ and 
	relaxation $\theta$ with $e\,\Phi_D^{\e} = (U+\Delta)/2$ and parameters $\Delta, U, \G$ as in 
	Fig. \ref{fig:ChargingDiagram}. It can be seen how the height of the cotunneling-mediated
	current step (a) and of the corresponding conductance peak (b) as well as their positions 
	depend on the relaxation rate. In the range of $0 \le \theta \lesssim \G^2 / \Delta$ (linear
	scale) the system is hardly affected by the relaxation. Between $\G^2/\Delta$ and $\Delta$
	(logarithmic scale) the height of current step and conductance peak first grow to a maximum 
	value at $\theta_0 \approx 0.58 \G$ (dashed lines in (a) and (b)) for increasing $\theta$, then
	decrease again and vanish before $\theta = \Delta$. Even faster relaxation (reciprocal scale) has
	no further effect. Figures (c) and (d) show cuts through (a) and (b), respectively, for five
	different values of $\theta$. Both the step in (a,c) and peak in (b,d) slightly shift towards
	higher absolute values of $e \VB$ for increasing rates between $\G^2 / \Delta$ and $\Delta$.}
\end{figure}

\section{Results} \label{sec:Results}

In this section we argue that the rich internal structure of the Coulomb diamond with its different overlapping regions and sub-regimes is responsible for the rather unexpected transport behavior the quantum dot shows in the presence of spin-relaxation. That is to say, the conductance peaks at the onset of sequential transport are, as stated above, maximally pronounced for a small finite relaxation rate. The peaks are situated close to the resonances with sequential transitions out of the excited state and therefore lie within the intermediate region. It turns out that the evolution of the peak height can be ascribed to the fact that in the core region of the Coulomb diamond the current $\tilde{I}(\theta)$ is much more sensitive to changes of the relaxation rate $\theta$ than it is in the shell. At small relaxation the current is diminished solely in the core region. Hence, the height of the current step is increased, while its width remains almost constant as compared to
zero relaxation. It follows that the resulting conductance signatures in the intermediate region grow with the relaxation rate as long as the latter stays below a level at which the current in the shell region is affected.

At first we describe general features of electron transport through the SLQD before we explain in detail how it depends on the relaxation parameter $\theta$. Fig. \ref{fig:ChargingDiagram}(b) shows a calculated charging diagram, i.e., the differential conductance $\tilde{G} = d \tilde{I} / d \VB$ against $e \VB$ and $e\,\Phi_D^{\e}$, of the cotunneling regime. The parameters are $\Delta = 45 k_{\text{B}} T$, $U = 225 k_{\text{B}} T$, $\G = 4.5 \times 10^{-3} k_{\text{B}} T$ and $\theta = \G / 2$. The Coulomb diamond is defined by pronounced red (dark gray) lines of high conductance, where one of the reservoir's electro-chemical potentials is close to resonance with the energy of a single-charge transition involving the ground state. Due to the Coulomb blockade the slightly weaker resonance lines of excited state transitions are not extended into the diamond. Instead, thin red (dark gray) lines appear in that part of the intermediate region, in which $\abs{e \VB} > \Delta$, corresponding to the yellow (light gray), not dotted area in Fig. \ref{fig:ChargingDiagram}(a). \cite{golovach:016601, schleser:206805} These are the signatures that mark the onset of cotunneling-mediated sequential transport out of the excited state. The light vertical shades along the lines $e \VB = \pm\Delta$ arise from the above-mentioned opening of inelastic transport channels. Though being actually invalid near the border of the Coulomb diamond and in its exterior, the approximate solution $\tilde{I}$ was used for Fig. \ref{fig:ChargingDiagram}(b) even in these regions, because its deviation from the systematic solution $I^{(1+2)}$ turned out---for the chosen set of parameters---to be too small to be visible. In general it may be necessary to use the systematic expansion for the border and outer region, which can be seamlessly connected to the approximate solution in the shell region given that $\G \b \ll 1$\cite{SmallDelta}.

\begin{figure}[htbp]

	\begin{minipage}{7.75cm}
		\includegraphics[scale=1]{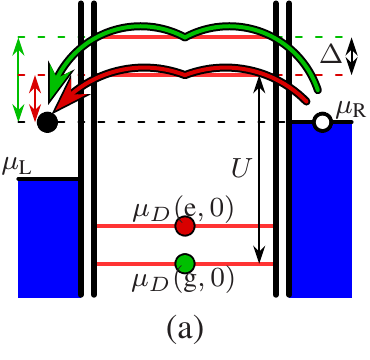}
		\hfill
		\includegraphics[scale=1]{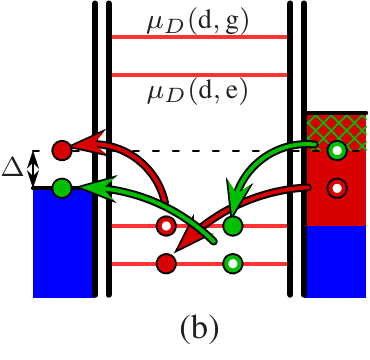}
	\end{minipage}

	\caption{\label{fig:CoreTunneling} (Color online).
	Position of the Fermi levels $\mu_{\text{L}}$ and $\mu_{\text{R}}$ of left and right
	reservoir (blue (very dark gray)) relative to the four energy differences 
	$\mu_D (\phi',\phi) = \e_{\phi'} - \e_{\phi} - e \Phi_D$ (light red (light gray) horizontal
	lines) between initial and finial state energy of the single-particle transitions (parameters
	$\Delta, U, \Phi_D^{\e}$ as in Fig. \ref{fig:ChargingDiagram}). In (a) bold green (light gray) 
	and red (dark grey) arrows represent elastic tunneling via virtual state $\QDD$ through the dot
	in ground and excited state, respectively, indicated by filled green (light gray) and red
	(dark gray) points ($e \VB = 1.5 \Delta$). Double pointed arrows on the left show for both 
	cases how much energy the tunneling electron must gain to get into the virtual
	state. (b) Coherent inelastic processes causing transitions $\QDg\to\QDe$ (green (light gray)
	arrows) and $\QDe\to\QDg$ (red (dark gray) arrows) are illustrated ($e \VB = 2 \Delta$).
	Electrons in the red (dark gray) colored (green (light gray) hatched) part of reservoir R can
	tunnel into the dot, if it's in initial state $\QDe$ ($\QDg$).}
\end{figure}

The dependence of current and conductance on the relaxation rate $\theta$ in the vicinity of the excited state resonance is shown in Fig. \ref{fig:DiffCondPeaks} part (a) and (b), respectively, (the range of bias values $e \VB$ is marked by the short horizontal line in Fig. \ref{fig:ChargingDiagram}(b)). The parameters $\Delta, U$, and $\G$ are the same as in Fig. \ref{fig:ChargingDiagram} and $e\,\Phi_D^{\e} = (U+\Delta)/2$. If the rate $\theta$ is smaller than $\G^2/\Delta$, thus well below the lowest second-order tunneling rate, the relaxation hardly affects $\tilde{I}$ and $\tilde{G}$, so that the system behaves as for $\theta = 0$. As the rates grow between $\G^2/\Delta$ and $\theta_0 \approx 0.58 \G$ they eventually become much larger than every cotunneling rate, while still being smaller than sequential rates of energetically allowed processes. Height and slope of the current step as well as the height of the conductance peak increase in this range to a maximum value at $\theta = \theta_0$, since the relaxation diminishes $\tilde{I}$ in the inelastic part of the core and low bias part of the intermediate region but leaves it almost unaltered in the sequential part of the shell region. A relaxation with rate $\theta_0 \leq \theta \leq \Delta$ is (roughly) as fast as or faster than sequential tunneling and while $\tilde{I}$ in this parameter range shows no further relaxation dependence in the core, it decreases in the shell region for growing $\theta$. As a result the current step and conductance peak decrease as well and vanish before $\theta = \Delta \gg \G$. For rates $\Delta \leq \theta \leq \infty$ transport properties do not depend on $\theta$. Obviously there exists an optimal relaxation rate $\theta_0$ for which the signatures of cotunneling-mediated sequential transport have maximal height and are considerably more pronounced than for $\theta = 0$. Also do the positions of the resonance signatures move to higher absolute values of $e \VB$, when $\theta$ is increased between $\G^2 / \Delta$ and $\Delta$. As $\theta_0$ depends on the ratio $\Delta/ U$, the coupling $\G$, and temperature $\b$ in a complicated way, we can give no simple estimation of its value in terms of the system parameters other than: $\theta_0 = c\, \G$ with $0.3 \lesssim c \lesssim 1.2$ for all parameter sets yielding reliable results. The height of the conductance peaks depends most strongly on $\theta$ throughout the whole range $(\G / \Delta)^{1/2} \lesssim \theta/\G \lesssim (\Delta / \G)^{1/2}$ with $\G \ll \Delta$ (see Fig. \ref{fig:DiffCondPeaks}(b)), while its relative variation for all values of $c$ amounts to only a few percent. Therefore, we content ourself with the statement, that $\theta_0$ corresponds to a rate that is roughly as large as $\G$ and very much larger than any cotunneling rate. In order to explain these observations, we examine how the tunneling processes that dominate the current in the relevant parts of the cotunneling regime are influenced by the relaxation. 

In the inelastic part of the core region the current is caused solely by elastic and inelastic cotunneling, which also dominates the occupation of the single-particle states. In particular, inelastic tunneling provides an occupation of the excited state of order 1 that doesn't depend on $\G$ and is reduced by the relaxation, as soon as the magnitude of $\theta$ becomes at least comparable to $W^{(2)}_{\text{e}\lto\text{g}}$. Since the $\tilde{\P}_{0,\text{d}}$ are much smaller than the single-particle occupations in the cotunneling regime, the current consists mainly of two contributions associated with cotunneling out of state $\QDg$ and $\QDe$, which are proportional to $\tilde{\P}_{\text{g}}$ and $\tilde{\P}_{\text{e}}$, respectively. Hence, the cotunneling based on processes with initial state $\QDg$ benefits from a change $d \tilde{\P}_{\text{g}} (\theta) \approx -d \tilde{\P}_{\text{e}} (\theta) > 0$ caused by relaxation, whereas the current with the dot being initially in state $\QDe$ is decreased by it. So in the core region the dependence of the current change $d \tilde{I}$ on $d \tilde{\P}_{\text{g}} (\theta)$ is given by $d \tilde{I} / d \tilde{\P}_{\text{g}} (\theta) \approx -e / \hbar \sum_{\phi'} \bigl( W^{L,(2)}_{\phi'\lto\text{g}} - W^{L,(2)}_{\phi'\lto\text{e}} \bigr)$. The sums on the rhs have the same sign, which is for both contributions specified by the direction of current flow and thus by the sign of $\VB$. For the inelastic part of the core one can establish the relation
\begin{equation}\label{eqn:CoreRateComparison}
	\Bigl| \sum_{\phi'} W^{L,(2)}_{\phi'\lto\text{e}} \Bigr| > 
	\Bigl| \sum_{\phi'} W^{L,(2)}_{\phi'\lto\text{g}} \Bigr|
\end{equation}
by looking at the energy dependence of elastic and inelastic processes. In the core, an electron that elastically tunnels through the dot with initial (and final) state $\ket{\chi = \text{g},\text{e}}$ via virtual intermediate state $\QDD$ has to overcome at least the energy difference $U + \d_{\chi,\text{g}}\Delta - e (\Phi_D^{\e} + \abs{\VB} / 2) > 0$, which is by $\Delta$ smaller for an initially excited dot than for one in the ground state (s. Fig. \ref{fig:CoreTunneling} (a)). The latter is also true for tunneling via virtual state $\ket{0}$, which can be seen analogously. As a consequence the rate for elastic cotunneling is smaller for $\chi = \text{g}$ than for $\chi = \text{e}$. 

For the inelastic processes a similar energy argument can be applied. Inelastic tunneling out of the ground into the excited state cannot set on before $e \abs{\VB} = \Delta$, because the energy $\Delta$, needed for the transition to take place, has to be provided by the reservoirs. In contrast, inelastic tunneling, causing the opposite transition, is always possible, because in this case the transition energy is provided by the dot. Hence, if $\VB$ and $\Phi_D^{\e}$ specify a point in the core region, for $\chi = \text{e}$ there are always more electrons available for inelastic processes compared to the case $\chi = \text{g}$ (s. Fig. \ref{fig:CoreTunneling} (b)). This results in a higher rate for inelastic tunneling through a dot in initial state $\QDe$ and immediately leads to Eqn. (\ref{eqn:CoreRateComparison}). Using this equation and the fact that $d \tilde{\P}_{\text{g}} (\theta)$ is a positive, monotonic function in $\theta$, for the core we can derive
\begin{equation} \label{eqn:CurrentRelaxDepCore}
    \frac{d \abs{\tilde{I}}}{d \theta} = \frac{e}{\hbar} \frac{d \tilde{\P}_{\text{g}}}{d \theta} 
        \biggl( \Bigl| \sum_{\phi'} W^{L,(2)}_{\phi'\lto\text{g}} \Bigr| - 
	\Bigl| \sum_{\phi'} W^{L,(2)}_{\phi'\lto\text{e}} \Bigr| \biggr) \le 0.
\end{equation}
Since in this region the single-particle occupations are determined mainly by cotunneling, they only depend on the relaxation, if $\theta$ is comparable to the second-order rates. Due to the factor $d \tilde{\P}_{\text{g}} / d \theta$ on the rhs of Eqn. (\ref{eqn:CurrentRelaxDepCore}), this dependence also holds for $\tilde{I}$, whose absolute value decreases for $\theta$ growing between $\G^2 / \Delta$ and $\G$ and is constant for slower or faster relaxation, respectively.

In the inelastic part of the shell region the maximal cotunneling-provided occupation of the excited state is by a factor $\G$ smaller than in the inelastic core, since it is reduced by sequential transport out of state $\QDe$, as long as $\theta \lesssim W^{\text{L},(1)}_{0,\text{d}\lto\text{e}} \approx \G$. When the relaxation becomes faster than sequential tunneling, $\tilde{\P}_{\text{e}}$ decreases and eventually goes to 0 for $\theta \gg \G$. The difference in size of sequential and cotunneling rates compensates for the reduction of the excited state occupation, so that the total current is higher in the shell compared to the core region. It consists of contributions associated with sequential tunneling out of states $\ket{\chi = 0,\text{e},\text{d}}$ and cotunneling out of the ground state.

When $\theta$ is not much larger than $\G$ these contributions are all of the same order of magnitude, which is, however, not the case for their response to increasing relaxation. Obviously, relaxation rates much higher than $\G$ completely depopulate the excited state and the current caused by tunneling out of $\QDe$ vanishes. Its relative change in magnitude compared to the case of low relaxation is therefore of order 1. For cotunneling out of state $\QDg$, on the other hand, the maximum relative change is
\begin{equation} \label{eqn:RelChangeGStateShell}
	\frac{\tilde{\P}_{\text{g}}(\theta \gg \G) - \tilde{\P}_{\text{g}}(\theta \ll \G)}
		{\tilde{\P}_{\text{g}}(\theta \ll \G)}
	= \frac{\mathcal{O}(\G)}{(1-\mathcal{O}(\G))}
	\approx \mathcal{O} (\G).
\end{equation}
Hence, if the relaxation increases, the gain in the cotunneling current associated with $\tilde{\P}_{\text{g}}$ cannot compensate for the simultaneous suppression of the sequential current proportional to $\tilde{\P}_{\text{e}}$, which results in a reduced total current. Similarly to the discussion of the core, however, it can be argued that in the shell, where the single-particle occupations are mainly determined by sequential processes, the total current can only show a considerable relaxation dependence, if $\theta$ is neither much smaller nor much larger than $\G$ or, equivalently, than the rates for sequential tunneling. As we stated above, the current dependence on $\theta$ smoothly crosses over between the core- and shell-like behavior in the intermediate region, so that both the current step and the conductance peak grow with $\theta$ between $\G^2 / \Delta$ and $\theta_0 \approx c\, \G$, while they decrease for $\theta > \theta_0$ and vanish before $\theta = \Delta$. The fact that the current becomes less sensitive to relaxation for higher values of $e \abs{\VB}$, showing a sharp step-like dependence in the intermediate region, also manifests in the slight shift of the position of the excited state resonances to higher absolute bias values.

\section{Summary} \label{sec:Summary}

In this paper we discussed Coulomb-blocked electron transport through a SLQD with spin-split level that is coupled to two non-magnetic, metallic leads. We used the real-time diagrammatic technique to systematically expand occupation probabilities and tunneling current up to the second-order in the strength $\G$ of the tunnel coupling, thereby including sequential and cotunneling into the transport calculations. Two properties were considered with respect to which the Coulomb blockade regime can be subdivided into parts that differ at least in one of them: the kind of tunneling processes contributing to the current (elastic, inelastic, sequential transport) and those determining the single-particle occupations (cotunneling in the core, sequential tunneling in the shell region). At or close to the borders between these sub-regimes, signatures of the dot's excitation spectrum appear in the current and differential conductance. With the focus on excited state signatures marking the onset of cotunneling-mediated sequential transport, we studied how the current is influenced by a phenomenologically introduced spin relaxation with rate $\theta$. It turned out that for a relaxation rate of about half the tunnel coupling the excited state resonances are maximally pronounced, being considerably larger than without relaxation, while in the limit of infinite $\theta$ the resonances completely vanish. We explained this behavior by a combination of two effects: (i) the current decreases monotonically with growing relaxation rates and (ii) the excited state occupation is in the cotunneling-dominated core and in the shell region only affected by a relaxation with rates in the range $\G^2/\Delta < \theta \lesssim c\, \G$ and $c\, \G \lesssim \theta < \Delta$ with $0.3 \lesssim c \lesssim 1.2$, respectively. 

This relaxation dependence of the current may illuminate why the resonance signatures measured in Ref. \onlinecite{schleser:206805} are relatively sharp compared to the ones that were calculated for $\theta = 0$. Furthermore it could provide means to directly influence the single-particle occupations in experiments and allows to facilitate measurements of excited state resonances by adjusting either the coupling $\G$ or the rate $\theta$. 

\section{Acknowledgments} \label{sec:Acknowledgments}

We thank J. K\"onig and M.R. Wegewijs for helpful discussions and acknowledge financial support from the Deutsche Forschungsgemeinschaft via SFB 508 "Quantum Materials".

\bibliographystyle{apsrev}


\begin{thebibliography}{48}
\bibitem[{Unc()}]{UnconventionalCoupling}
\bibinfo{note}{Thus, we define the coupling strength as $(2\pi)^{-1}$ times the
  rate for sequential tunneling (obtained by Fermi's golden rule).}
\bibitem[{Sta()}]{StationaryLimit}
\bibinfo{note}{In the stationary state, we can then w.l.o.g. set $t=0$.}
\bibitem[{Dia()}]{DiagramStructure}
\bibinfo{note}{This can be seen by looking at the structure of the diagrams
  that constitute propagator $\op{\Pi}$.}
\bibitem[{Rel()}]{RelaxDefWeymann}
\bibinfo{note}{For a fermion bath with $\mu_{\text{bath}} = 0$ these rates lead
  to the same relaxation terms as in Eq. (2) of Ref.
  \onlinecite{weymann:205309}, if we identify $\theta = 2 \hbar/
  \tau_{\text{sf}}$.}
\bibitem[{Sma()}]{SmallDelta}
\bibinfo{note}{Of course, if $\Delta$ is too small to lead to a sufficiently
  broad shell region, no (smooth) crossover can be found.}
\end{thebibliography}

\end{document}